      [1999/12/01 v1.4c Il Nuovo Cimento]
\documentclass{cimento}

\usepackage{graphicx}  

\title{GRIPS - The potential of a future MeV survey}
\author{J.~Greiner\from{ins:x},
G.~Kanbach\from{ins:x},
K.~Mannheim\from{ins:y}\\
        \atque
the GRIPS collaboration\thanks{www.grips-mission.eu}}
\instlist{\inst{ins:x} Max-Planck Institut f\"ur extraterrestrische Physik,
       85740 Garching, Giessenbachstr. 1, Germany
  \inst{ins:y} W\"urzburg Univ., Am Hubland, D-97074 W\"urzburg, Germany
}
\PACSes{\PACSit{00.00}{26.50.+x, 95.30.Gv, 95.55.Ka, 95.85.Pw}
\PACSit{---.---}{\ldots}}
\begin{document}

\maketitle

\begin{abstract}
We describe the potential of GRIPS, a future MeV mission.
The Gamma-Ray Imaging, Polarimetry and Spectroscopy (``GRIPS'') concept
combines a Compton and pair telescope, and will be a very sensitive 
polarimeter. GRIPS would perform a continuously scanning 
all-sky survey from 200 keV to 80 MeV achieving a sensitivity which is 
better by a factor of 40  compared
to the previous missions in this energy range.
\end{abstract}

\section{Introduction: The MeV gap}

The photon energy range between hard X-rays of 0.2\,MeV and$\gamma$-rays of 
80\,MeV covers the prime range of nuclear excitation      
and binding energies. Many high energy sources have their peak emissivity
in this regime, which therefore is as important for high-energy
astronomy as optical astronomy for phenomena related to atomic physics.
In addition, it includes the energy scale of the electron and pion rest mass.
The ``MeV-gap'' in current instrument sensitivity stretches
exactly over this range (Fig.~1). The GRIPS mission \cite{gik09} 
will improve the 
sensitivity in this gap by a factor of 40 compared to previous missions.
Therefore, the GRIPS all-sky survey with $\gamma$-ray imaging, 
polarimetry, and spectroscopy holds a high promise of new discoveries 
and of precision diagnostics of primary high-energy processes.

The scientific scope of the mission
is centered on the themes ``The Evolving Violent Universe'' 
and ``Matter under extreme conditions'' of the Cosmic Vision strategic plan, 
pertaining to the astrophysics of the most extreme objects
in the Universe in which the plasma becomes relativistic, nuclear interactions 
and radioactive decays take place, and 
where particle acceleration plays a major role in the energy budget. 

\begin{figure}[th]
\centering
   \includegraphics[width=0.7\textwidth]{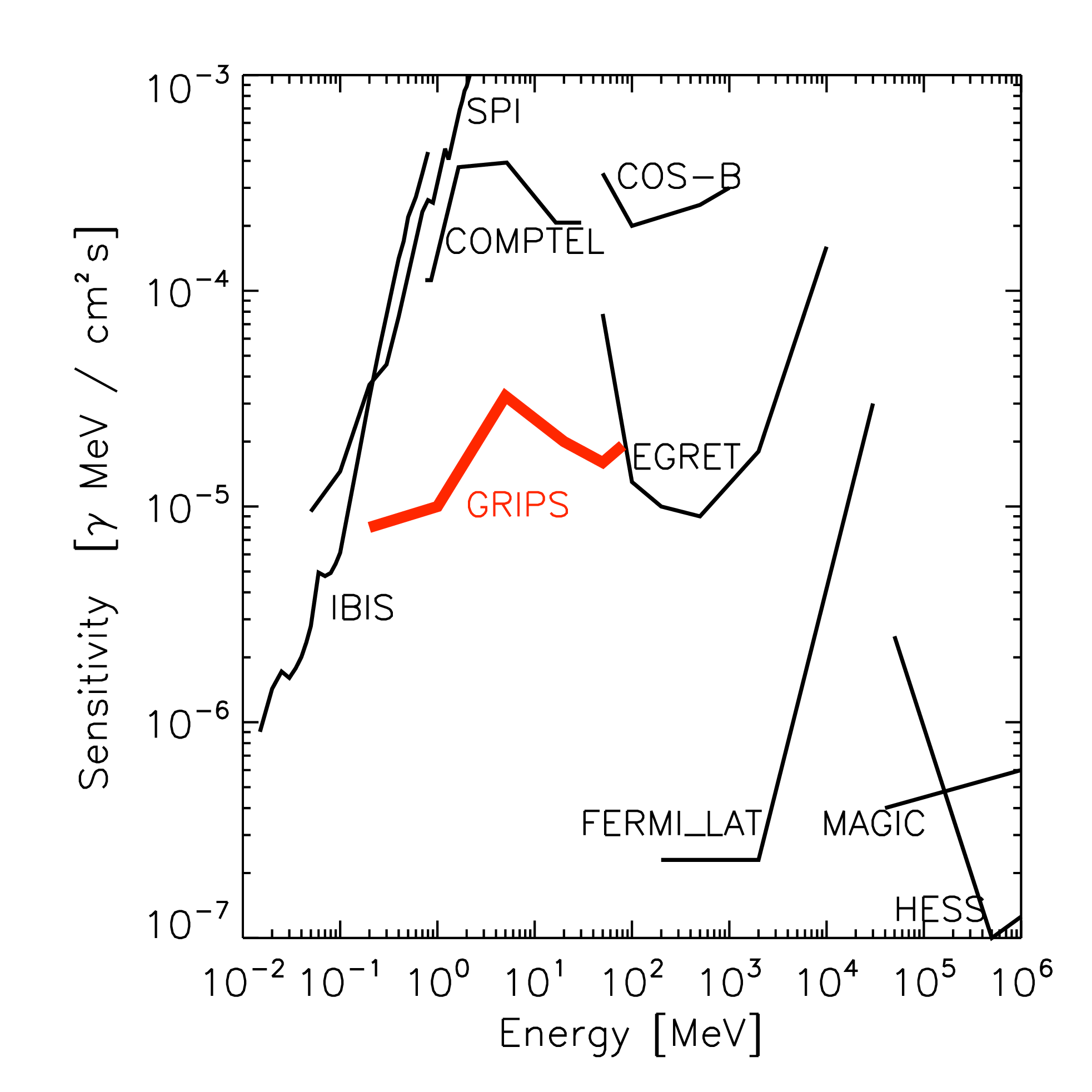}
\caption{GRIPS will allow a major sensitivity improvement
in an energy range (between hard X-rays and GeV $\gamma$-rays) which 
has been poorly explored, yet holds unique information for a wide range 
of astrophysical questions. The curves are for an exposure of 10$^6$ sec,
$\Delta E = E$, and an $E^{-2}$ spectrum.}
\label{fig_sensitivity}
\end{figure}

GRIPS will open the astronomical window to the bizarre and highly variable 
0.2--80 MeV sky, to investigate fascinating cosmic objects 
such as $\gamma$-ray bursts, blazars, supernovae and their remnants, 
accreting binaries with white dwarfs, neutron stars or black holes
often including relativistic jets,
pulsars and magnetars, and the often peculiar cosmic gas in their 
surroundings. 
Many of these objects show MeV-peaked spectral energy distributions or 
characteristic spectral lines; we target such primary emission to understand 
the astrophysics of these sources. 

\begin{figure}[th] 
\centering
\hspace{1.8mm}\includegraphics[width=0.60\columnwidth]{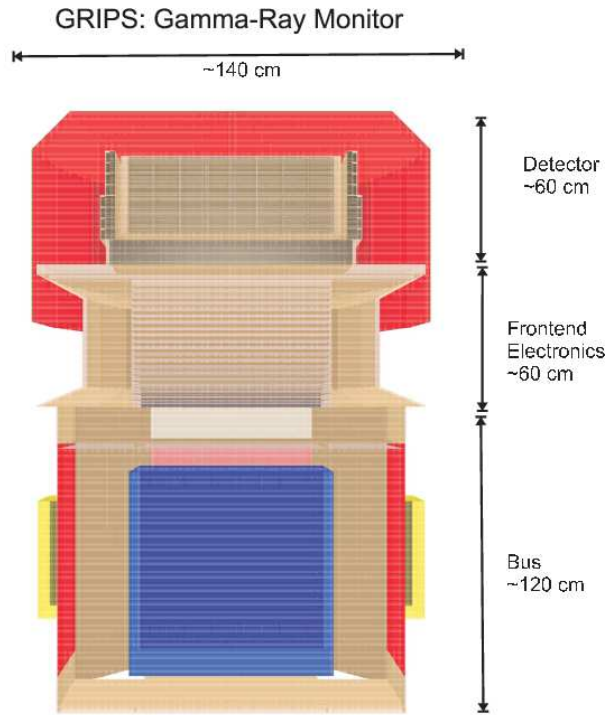} 
\caption[Size of the Gamma-Ray Monitor]{Expected size and assembly 
of the Gamma-Ray Monitor (top) and related electronics (center) on a  
generic satellite bus (bottom). 
\label{3D}} 
\vspace{-0.6cm}
\end{figure}

\section{The Mission}

GRIPS will combine a Compton and pair telescope based on the latest 
developments in nuclear and high-energy physics laboratories (Fig. \ref{3D}). 
Modern 3D position-sensitive and space-proven detectors with advanced 
(nanosecond level) readout technology will ensure unprecedented background 
rejection capability \cite{zog08}
and thus guarantee the above sensitivity leap. Taking 
advantage of the Compton-scattering physics, GRIPS will also be a very 
sensitive polarimeter \cite{gre09}.  
The energy resolution of 3\% at photon energies of 
1 MeV renders the gamma-ray telescope ideal for the study of broadened emission
lines from explosive sources such as supernovae; and with the extended 
spectroscopic performance throughout the entire nuclear energy range we will 
be armed for pioneering astrophysical studies of nuclear excitation and 
resonance absorption lines.  
The limitation in imaging resolution which is intrinsic to the detection 
physics in the MeV band will be compensated by detecting secondary emission 
from the same sources with auxiliary X-ray and NIR telescopes with their 
sub-arcmin angular resolutions.

\begin{figure}[th]
\includegraphics[width=0.45\columnwidth]{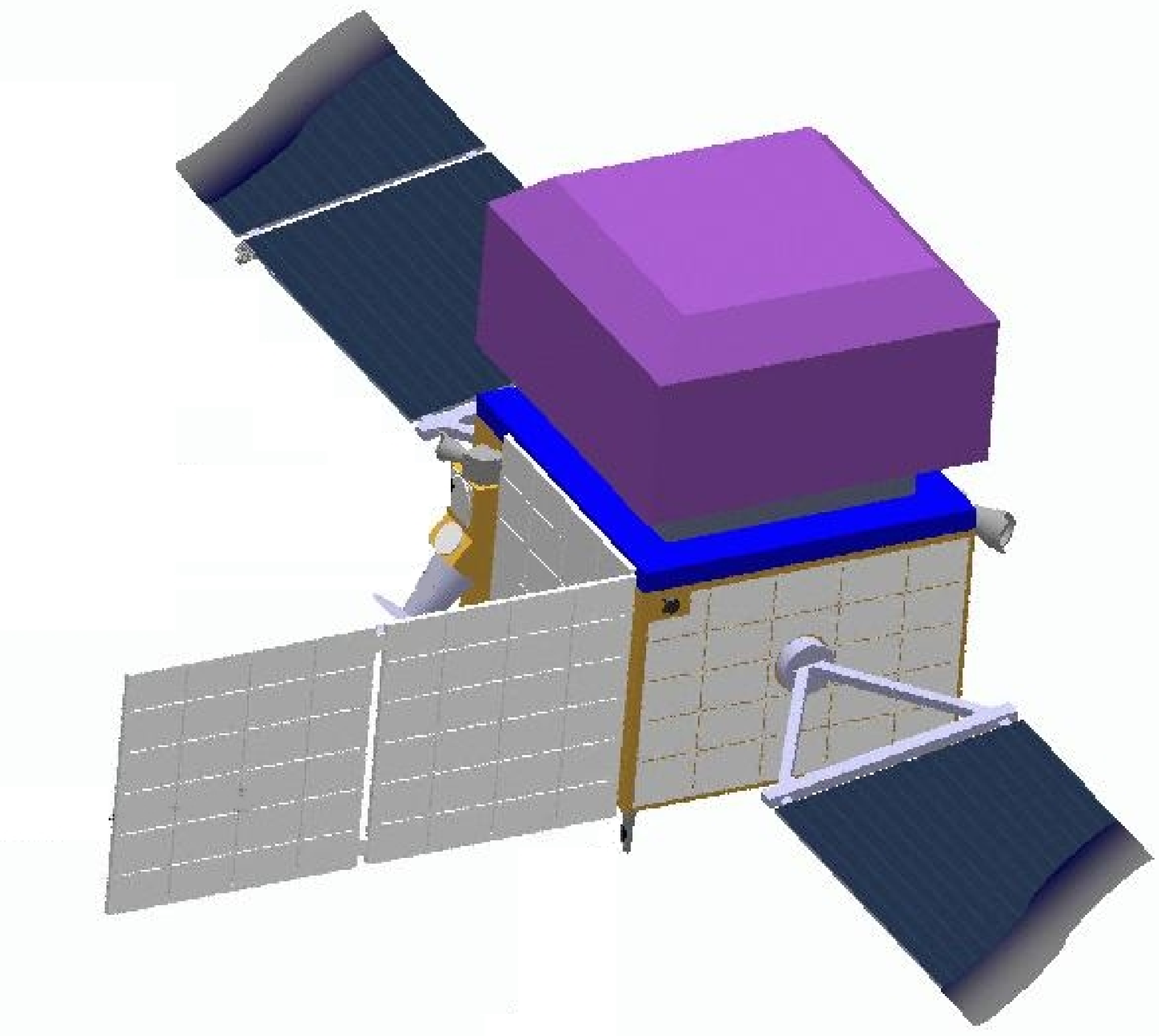}
\hfill
\includegraphics[width=0.45\columnwidth]{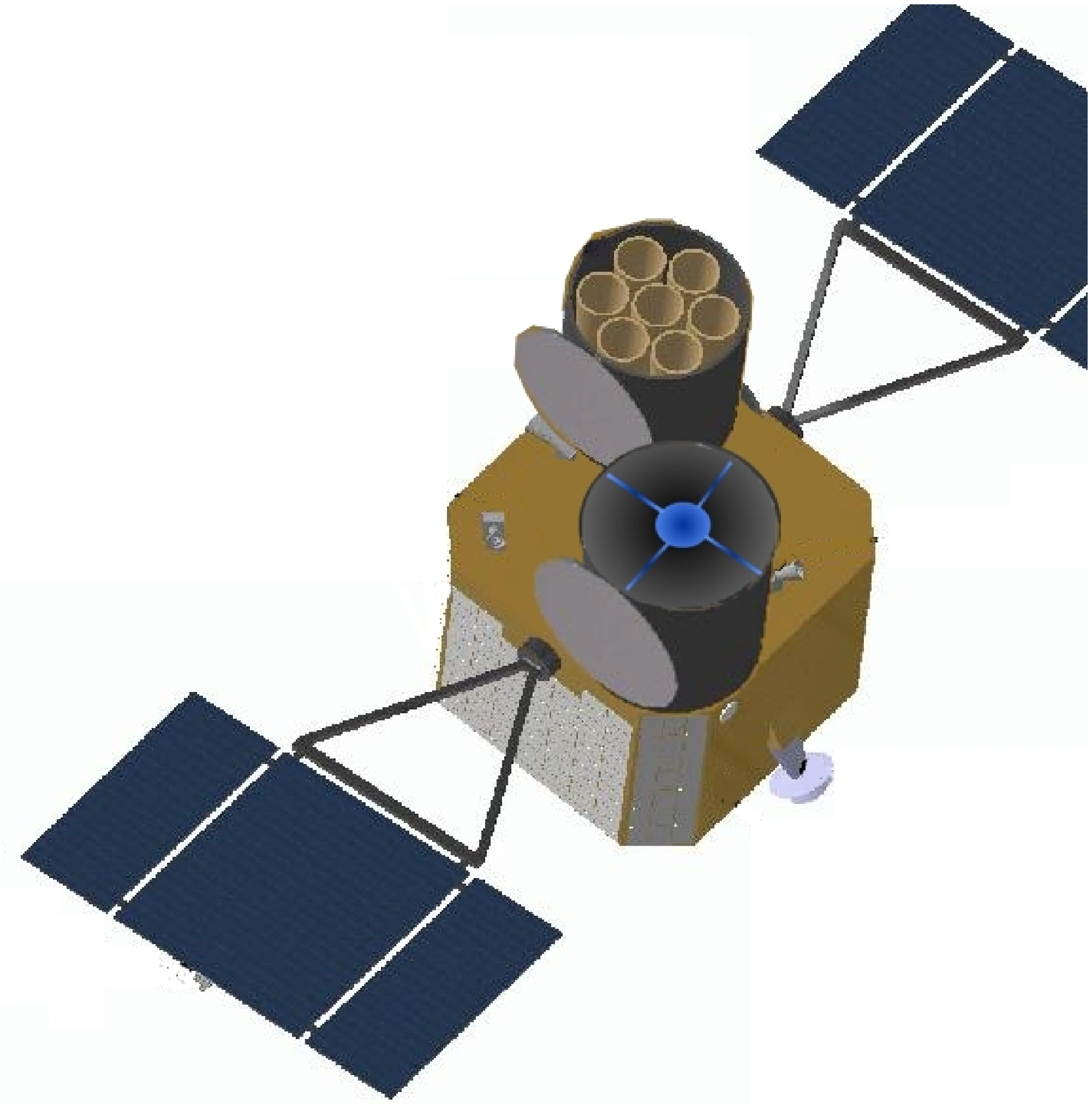}
\caption[GRIPS_2-sat]{GRIPS configuration in the two-satellite option, 
where the GRM is on one satellite (left), and XRM and IRT on the other
(right). The GRM satellite would just do the zenith scanning all-sky survey,
while XRM/IRT would re-point with the whole (second) satellite to GRBs,
similar to {\it Swift}. }
\label{fig_2sat}
\end{figure}

GRIPS should be launched into a low-altitude,
equatorial orbit (LEO) to minimize the background. 
GRIPS consists of two satellites,
flying in a ``close-pair'' configuration (Fig. \ref{fig_2sat}): 
one satellite with the 
Gamma-Ray Monitor (GRM), the other with an X-Ray Monitor XRM and an
Infra-Red Telescope (IRT). Both satellites should be 3-axis stabilized.
The gamma-ray telescope/satellite of GRIPS will be continuously 
pointing at the zenith, thus monitoring ca. 80\% of the sky over 
each orbit for transient events, including gamma-ray bursts. 
The X-ray/infrared telescope satellite should have the capability of 
autonomously slewing for follow-up observations of gamma-ray bursts. 
As added value, the mission will deliver positions and fluxes of
transient alerts to the community. For gamma-ray bursts, bursts
in the redshift range $7<z<35$ will be recognized within minutes by IRT,
and this information also transmitted to ground in near-real-time.
We anticipate a lifetime of 10~years.  The prime instrument will use 500 kg 
of LaBr$_3$ scintillator crystals, and the total satellite weight is about
5~tons. The launcher should be a Soyuz-Fregate rocket launched from Kourou. 
The spacecraft busses should provide 1.8/1.1~kW power and 40~Mbps telemetry 
bandwidth.

\section{Scientific Themes}

{\bf Gamma-Ray Bursts and First Stars:}
Unrivaled by any other method, the detection of highly penetrating 
$\gamma$-rays from cosmological $\gamma$-ray bursts 
will shed  light on the first massive stars and galaxies which formed
during the dark ages of the early Universe.
With its energy coverage up to 80 MeV, GRIPS will firmly establish 
the high energy component seen in addition to the canonical
Band funtion in one {\it CGRO}/EGRET ($>$10 MeV) and  
one {\it Fermi}/LAT burst ($>$100\,MeV) in  much  larger numbers,  
and characterize its origin through polarisation signatures. 
GRIPS will measure the degree of polarisation of the
prompt $\gamma$-ray burst emission to a few percent accuracy for more
than 10\% of the detected GRBs, and securely
measure how the degree of polarisation varies with energy
and/or time over the full burst duration for dozens of bright GRBs.
Also, the delay of GeV photons relative to emission at $\sim$ hundred keV,
observed in a few GRBs with {\it Fermi}/LAT, manifests itself already
at MeV energies in {\it Fermi}/GBM, and will thus be a science target
for GRIPS. These observations enable a clear identification of the prompt 
GRB emission processes, and determine the role played by magnetic fields.

GRIPS will detect about 650 GRBs yr$^{-1}$, a large fraction of of these 
at high-redshift ($\sim$30 GRBs yr$^{-1}$ at $z>5$, and $\sim$22 GRBs
at $z>10$). The  7-channel
near-infrared telescope (IRT) will improve  the localization to the 
required arcsecond
level and  will determine  photometric redshifts for  the bulk  of the
most distance ($z>7$) sources. 
It will allow to measure the incidence of gas and metals through X-ray 
absorption
spectroscopy and line-of-sight properties by enabling NIR spectroscopy
with {\it JWST}. 
 
If  the  GRB  environments contain total hydrogen column densities of 
10$^{25}$\,cm$^{-2}$, or higher, GRIPS holds the promise of measuring  
redshifts directly from the $\gamma$-ray spectrum via nuclear resonances, 
and will be sensitive to do so beyond z$\sim$13.

GRIPS will also detect a handful of short GRBs at $z < 0.1$, enabling a
potential discovery of correlated gravitational-wave and/or neutrino signal.

\begin{figure}[t] 
\includegraphics[width=0.75\columnwidth]{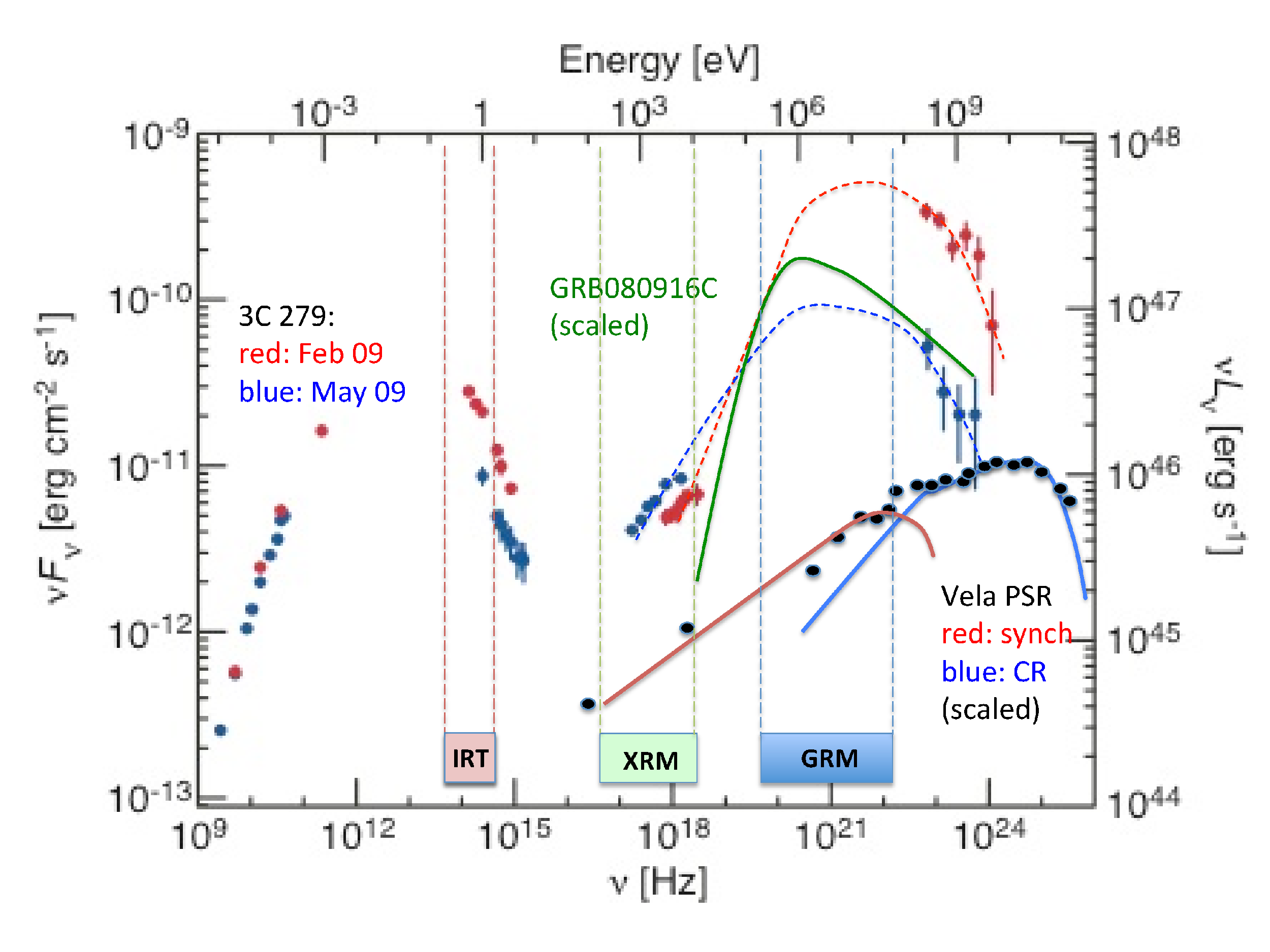}
\hfill\parbox[t]{3.cm}{\vspace*{-5.5cm}\caption{Not only GRBs, 
but also blazar SEDs peak in the MeV range,
and pulsars turn over from their maximum in the {\it Fermi} band.
The combined $\gamma$-, X-ray and near-infrared coverage of blazars
covers both emission components simultaneously. }} 
\label{fig_blazar-sed} 
\end{figure}

{\bf Blazars:}
GRIPS will catalogue about 2000 blazars, probing blazar evolution to large
redshifts. These observations will pinpoint the most massive halos at large
redshifts, thus severely constraining models of structure evolution.
This large sample will establish their (evolving) luminosity
function and thus determine the fractional contribution of blazars to the
diffuse extragalactic background. GRIPS is expected to detect $\sim$10
blazars at z $>$ 8.
Studies of the nonthermal radiation mechanisms will be supported
through spectro-polarimetric measurements. The link between the inner 
accretion disk and the jet can be probed with correlated variability 
from the thermal to the nonthermal regime, using GRIPS auxiliary 
instruments. This will localize the region of high-energy emission.

{\bf Supernovae and Nucleosynthesis:}
The primary energy source of supernova (SN) light is radioactive decay. 
The first direct measurement of the nickel and cobalt decay in Type~Ia SNe 
will pin down their explosion physics and
disentangle their progenitor channels. This will impact the luminosity 
calibration of Type~Ia SNe that serve as standard candles in cosmology.
The otherwise unobtainable direct measurement of the inner ejecta and the 
explosive nucleosynthesis of core
collapse supernovae will allow to establish a 
physical model for these important terminal stages of massive-star evolution.
Explosion asymmetries 
and the links to long GRBs are important aspects herein.
The fraction of nearby
pair-instability supernovae from very massive stars will be unambiguously 
identified through their copious radioactivity emission.
All hese observations will be crucial for complementing  
neutrino and gravitational wave measurements, and for our understanding of 
cosmic chemical evolution.

{\bf Cosmic Rays:}
Nuclear de-excitation lines of abundant isotopes like $^{12}$C and $^{16}$O, 
the hadronic fingerprints of cosmic-ray acceleration, 
are expected to be discovered with GRIPS.
Understanding the relative importance of leptonic and hadronic processes, 
and the role of cosmic rays in heating and ionizing molecular clouds 
will boost our understanding of both relativistic-particle acceleration and 
the cycle of matter.

{\bf Magnetars:}
The detection of instabilities in the supercritical magnetospheres
of magnetars, which are expected to lead to few-hundred keV to possibly 
MeV-peaked emission, 
will explore white territory on the field of plasma physics.

{\bf Annihilation of Positrons:} 
GRIPS will probe positron escape from candidate sources along the galactic plane
through annihilation $\gamma$-rays in their vicinity. For several microquasars
and pulsars, point-source like appearance is expected if the local annihilation
fraction $f_{local}$ exceeds 10\% (I$_{\gamma}\sim10^{-2}\cdot f_{local}$
ph~cm$^{-2}$s$^{-1}$).
GRIPS will enable the cross-correlation of annihilation $\gamma$-ray images with
candidate source distributions, such as $^{26}$Al and Galactic diffuse
emission above MeV energies (where it is dominated by cosmic-ray interactions
with the ISM) [both also measured with GRIPS at superior quality], 
point sources derived from {\it INTEGRAL}, {\it Swift}, {\it Fermi}, and 
{\it H.E.S.S./MAGIC/VERITAS/CTA} measurements,
and with candidate dark-matter related emission profiles.
GRIPS will deepen the presently best {\it INTEGRAL} sky image by at least an 
order of
magnitude in flux, at similar angular resolution. Comparing Galactic-disk and
-bulge emission, limits on dark-matter produced annihilation emission will 
constrain decay
channels from neutralino annihilation in the gravitational field of our Galaxy.
GRIPS will also perform sensitive searches for $\gamma$-ray signatures of dark
matter for nearby dwarf galaxies.

{\bf Solar Flares:}
Solar flares will be a natural by-product of the continuous sky survey carried
out by GRIPS since the Sun passes regularly through its field of view. 
Gamma-rays
in the MeV regime provide the means to directly probe particle acceleration and
matter interactions in these magnetised, non-thermal plasmas. Polarisation
measurements are of great value for disentangling these dynamic processes.

\section{Conclusion}

We have the following burning questions: 
{\it How do stars explode?}
{\it What is the structure of massive-star interiors and of compact stars?}
{\it How are cosmic isotopes created and distributed?}
{\it How does cosmic-ray acceleration work?}
{\it How is accretion linked with jets?}
Answering these questions will provide the basis to understand the larger 
astrophysical scales,
like the interstellar medium evolution in galaxies, the supernova-fed gas 
in galaxy clusters, 
and the cosmic metallicity evolution.  This {\it bottom-up}
approach will also help in reaching other ambitious goals of the 
{\it Cosmic Vision} plan which addresses
the challenging fundamental physics near event horizons, and cosmological 
questions which may carry us beyond standard astrophysics as we know it today. 
GRIPS will be an extraordinary tool to advance the study of the 
{\it nonthermal and violent Universe}. 

The GRIPS mission would provide the data to answer key questions of high-energy 
astrophysics. Moreover, the all-sky survey with an expected number of more 
than 2000 sources, many of them new, will at the same time
serve a diversity of communities for the astronomical exploration of so-far 
unidentified X/$\gamma$-ray sources and of new phenomena. 
This will strengthen ESA's outstanding heritage in pioneering space research. 
The delivery of triggers on bursting sources of high-energy emission will 
amplify the scientific impact of GRIPS across fields and communities.
As the 2010 Decadal Survey Report of the US Academy of Science puts it, 
``Astronomy is still as
much based on discovery as it is on predetermined measurements."

\end{document}